\documentstyle[12pt,aaspp4]{article}
\tighten

\lefthead{Band and Hartmann}
\righthead{GRB No Host Problem}

\begin{document}

\title{A STATISTICAL TREATMENT OF THE GAMMA-RAY BURST 
``NO HOST GALAXY'' PROBLEM: \\ I. METHODOLOGY}
\author{David L. Band}
\affil{CASS 0424, University of California, San Diego, La Jolla, CA  
92093; dband@ucsd.edu}
\author{Dieter H. Hartmann}
\affil{Department of Physics and Astronomy, Kinard Laboratory of
Physics, Clemson University, Clemson, SC 29634;
hartmann@grb.phys.clemson.edu} 
%
\centerline{\it Received 1997 July 31; accepted 1997 September 9}
\centerline{To appear in the 1998 February 1 (volume {\bf 493})}
\centerline{issue of {\it The Astrophysical Journal}}
\begin{abstract}
If gamma-ray bursts originate in galaxies at cosmological distances,
the host galaxy should be detected if a burst error box is searched
deep enough; are the host galaxies present?  We present and implement
a statistical methodology which evaluates whether the observed galaxy
detections in a burst's error box are consistent with the presence of
the host galaxy, or whether all the detections can be attributed to
unrelated background galaxies.  This methodology requires the
model-dependent distribution of host galaxy fluxes.  While our
methodology was derived for galaxies in burst error boxes, it can be
applied to other candidate host objects (e.g., active galaxies) and to
other types of error boxes. As examples, we apply this methodology to
two published studies of burst error boxes.  We find that the nine
error boxes observed by Larson \& McLean (1997) are too large to
discriminate between the presence or absence of host galaxies, while
the absence of bright galaxies in the four significantly smaller error
boxes observed by {\it HST} (Schaefer et al. 1997) does confirm that
there is a ``no-host galaxy'' problem within the ``minimal'' host
galaxy model. 
\end{abstract}
\keywords{gamma-rays: bursts---methods: statistical}
\section{INTRODUCTION}
If gamma-ray bursts originate at cosmological distances then they most
likely occur in (or near) galaxies; are the host galaxies present in
burst error boxes? There have been various claims as to whether the
error boxes which have been searched in the optical band contain
galaxies bright enough to be the expected hosts of the burst sources. 
The question is whether the observations---the galaxies observed above
the detection threshold---are consistent with the presence of the
expected host galaxy when unrelated ``background'' galaxies are also
be present. Here we present and implement a methodology to evaluate
this question. 


The ``no-host galaxy'' problem was first raised by Schaefer (1992) who
presented a compendium of brightness upper limits for galaxies in the
error boxes of 8 classical bursts, the soft gamma-repeater associated
with the 1979 March~5 event and 4 optical transients.  For each error
box Schaefer calculated the distance to a galaxy with the luminosity
of M31 if the galaxy were as bright as the detection threshold for
that error box.  The distances were typically a few Gpc (a Gpc
corresponds to $z=0.25$ for $H_0=75$~km~s$^{-1}$~Mpc$^{-1}$),
requiring isotropic radiated energies greater than $10^{53}$~erg in
some cases. Note that Schaefer did not claim that there are no
galaxies in the error boxes, just that there are no bright galaxies. 
Indeed, he used the brightest object in his field as the upper limit
for the brightest galaxy. 

Subsequently Fenimore et al. (1993) analyzed the PVO-BATSE cumulative
burst intensity distribution under the assumption that bursts are
cosmological, and assigned distances to the 8 classical bursts in
Schaefer's sample based on the bursts' intensities. Using the distance
estimate and Schaefer's brightness limit for each error box, Fenimore
et al. calculated first the upper limit to the host galaxy's
luminosity and then the fraction of the host galaxy luminosity
function which is fainter than this limit. They derived the host
galaxy luminosity function as the normal galaxy luminosity function
weighted by the luminosity since the number of potential burst sources
in a galaxy presumably scales with the number of stars in the galaxy,
and therefore with the luminosity (for a constant mass-to-light
ratio). If only host galaxies are present, the fraction of the host
galaxy luminosity function less than the observed host galaxy's
luminosity should be distributed uniformly between 0 and 1, with an
average of 1/2.  Indeed, the analysis of Fenimore et al. gives an
average value of 0.44$\pm0.10$ for the upper limits calculated from
Schaefer's compendium. However, the average value of this fraction for
the host galaxies' actual luminosities is undoubtedly smaller (at
least some of the host galaxies must be fainter than the upper
limits), and Fenimore et al. concluded that the observations were only
marginally consistent with bursts occurring in galaxies. 

Vrba, Hartmann, \& Jennings (1995) monitored the error boxes of 7
classical bursts and one optical transient for 5 years; many of the
burst error boxes were included in the compendium of Schaefer (1992).
Vrba et al. searched for, but did not find, unusual objects which
varied or had bizarre colors.  Many galaxies were identified based on
morphology (for $V<21.6$) and color (for fainter objects)---the number
of galaxies was consistent with galaxy counts---but whether the error
boxes contained a galaxy bright enough to be the expected host galaxy
was not considered. 

Larson \& McLean (1997) observed in the infrared nine of the smallest
error boxes of classical bursts localized by the third Interplanetary
Network (Larson, McLean \& Becklin 1996 presented a preliminary report
on six error boxes); these error boxes are typically $\sim8$
arcmin$^2$.  In or near all but one error box they found at least one
bright galaxy (K$\le 15.5$). The fraction of the host galaxy
luminosity function fainter than the brightest galaxy in each error
box, the statistic introduced by Fenimore et al. (1993), has an
average of 0.47$\pm0.10$, consistent with the value 0.5 which is
expected if only host galaxies are present (Larson 1997).  Larson \&
McLean (1997) recognize that the error boxes are too large to discern
between the host galaxy and unrelated background galaxies. However,
they report that the surface density of bright galaxies is
approximately a factor of two larger than the average, which they
interpret as possible evidence for clustering around the host galaxy. 

Schaefer et al. (1997) searched the error boxes of 5 classical bursts
with the {\it Hubble Space Telescope (HST)}; 4 of the error boxes are
small (of order $\sim1$ arcmin$^2$) and 3 are in the compendium of
Schaefer (1992).  This study also looked for, but did not find,
unusual objects with ultraviolet excesses, variability, parallax or
proper motion.  Galaxies are present, but faint.  Following Fenimore
et al. (1993), Schaefer et al. estimated the distance to the host
galaxies using the bursts' peak flux, with the distance scale
determined from the burst cumulative intensity distribution.  The
brightest object in each field which could be a galaxy sets the upper
limit on the host galaxy's brightness.  An upper limit on the host
galaxy's luminosity is derived from this observed brightness upper
limit and the estimated distance; for the 4 small error boxes the
luminosity upper limits are 10-100 times smaller than the luminosity
of an $L_*$ galaxy. 

Thus the issue is not whether there are galaxies in burst error boxes,
but whether the observed galaxies are likely host galaxy candidates.
Galaxies may be present, but they may be fainter than the expected
host galaxy brightness.  Alternatively, the error box may be so large
that the host galaxy is only one of the many expected unrelated
background galaxies.  The statistical analyses of the observations
have thus far treated the brightest object within the error box as the
host galaxy's brightness upper limit, even though many galaxies were
detected. These analyses have not considered the size of the error box
or the effect of the expected background galaxies.  In addition, most
analyses of error boxes treat the region within a certain confidence
contour, typically 99\%, as equally likely, whereas in reality the
location probability density peaks within the error box. 

Therefore, to evaluate whether the host galaxy is present we have
developed and implemented a methodology which considers the location
probability density (and thus the size of the error box), all the
detected galaxies, and the presence of background galaxies.  While
this methodology was derived to determine whether host galaxies are
present in burst error boxes, it can be applied to other candidate
host objects, such as active galaxies (Luginbuhl et al. 1996).  Of
course, the methodology can also be used in other astrophysical
contexts where a counterpart in one wavelength band is sought for a
source observed in a second band. Although the methodology is derived
within a Bayesian framework, the resulting Bayesian ``odds ratio'' can
be understood intuitively, and thus can be treated as a non-Bayesian
statistic.  The ultimate purpose of the odds ratio is to determine
whether host galaxies are present, but it can also be used to
determine which error boxes will have the power to answer this
question.  We use the standard notation where $p(a \,|\,b)$ is the
conditional probability of proposition $a$ given proposition $b$.
Propositions are simple statements, the validity of which may or may
not be in question. 

This methodology deals with error boxes which have a finite size and
assumes that bursts occur within visible host galaxies.  If the
optical transients in the Beppo-SAX error boxes do indicate the
locations of the GRB~970228 and GRB~970508 bursts, then these bursts
(and subsequent similarly localized bursts) can be treated as having
much smaller error boxes than previous bursts.  Our methodology can be
applied to these bursts by using as the error box the region within
which a galaxy would be acceptable as a host.  The GRB~970228 optical
transient appears to sit on a slightly extended source which may or
may not be a galaxy (Djorgovski et al. 1997 report that an R=25.5
source is still present half a year after the burst). No extended
source has been associated with the GRB~970508 optical transient. 
This has revived suggestions that the burst source is expelled from
the host galaxy (e.g., Lipunov et al. 1995; Bloom, Tanvir, \& Wijers
1997). Source ejection can be treated within our methodology by
expanding a burst's error box by the (model dependent) angular
distance the burster would have traveled before bursting. 

Any analysis of the host galaxy issue depends on the expected host
galaxy distribution.  The examples presented here, the analyses
published elsewhere and indeed most studies of the possible
cosmological properties of burst ensembles use a ``minimal''
cosmological model.  In this model bursts are assumed to be standard
candles which do not evolve in comoving density or luminosity.
Modeling the intensity distribution gives a unique mapping between a
burst's intensity and its distance.  Bursts occur in galaxies at a
rate proportional to a galaxy's mass and (assuming a constant
mass-to-light ratio for all galaxies) therefore luminosity.  This is 
most likely overly simplistic and in the future we will relax various 
assumptions of this minimal model.  However here, to demonstrate our 
methodology, we will test the minimal cosmological model.

In \S 2.1 we develop the methodology, which is dependent on the model
host galaxy distribution, as discussed in \S 2.2.  The ability of the
observations of a given error box to discriminate between the presence
or absence of the host galaxy can be evaluated using this methodology
(\S 2.3).  Using the minimal cosmological model, we apply this
methodology in \S 3 to the K-band observations of Larson \& McLean
(1997) and the {\it HST} observations of Schaefer et al. (1997).
Finally in \S 4 we discuss important issues raised by our methodology.
\section{METHODOLOGY}
\subsection{Likelihood Ratio}
The basic strategy for evaluating whether the host galaxy may be
present uses the three-dimensional space consisting of the two sky
coordinates $\Omega$ and the optical flux $f$ at each position. The
detection threshold $f_{\rm lim}$ may vary over the searched region;
for example, a bright star reduces the detectability of faint galaxies
in the star's immediate vicinity. The searched region may not include
the entire burst error box and may extend beyond it. Assume that $n_d$
galaxies are detected, each with a flux $f_i$ located at $\Omega_i$.
An observed galaxy is either the host galaxy or an unrelated
background galaxy. The searched portion of the flux-position space
should be visualized as little bins, most of which are empty;
expressions are calculated for finite-sized bins which are
subsequently reduced to infinitesimal dimensions. The probability for
obtaining the observations is the product of the probabilities for the
observed detection or nondetection in each bin. 

Let the distribution of background galaxies be $dN = \phi(f) \, df
d\Omega$ where we assume that these galaxies are distributed uniformly
without any clustering. If it exists, the burster host galaxy is drawn
from $\Psi(f)$ which must be normalized to 1 since there can only be
one host galaxy per error box.  The burst localization results in a
probability density $\rho(\Omega)$ for the burst's position on the
sky; $\rho$ is also normalized to 1.  Therefore, the probability that
the host galaxy is at position $\Omega$ with flux $f$ is $p(f,\Omega)
= \Psi(f) \rho(\Omega)$. 

The three dimensional flux-position space is broken into bins with
dimensions $\Delta \Omega$ and $\Delta f$.  The probability of finding
a background galaxy in a bin is governed by Poisson statistics; the
expected number of galaxies in a bin centered on $f_j$ is $n_j =
\phi(f_j) \Delta f \Delta \Omega$. The detection probabilities are 
\begin{eqnarray}
p_j(n=0) =& e^{-n_j} &= e^{- \phi(f_j)\Delta f \Delta\Omega} \nonumber \\
p_j(n=1) =& n_j e^{-n_j} &= \phi(f_j) \Delta f \Delta \Omega 
   e^{- \phi(f_j) \Delta f \Delta \Omega} \quad . 
\end{eqnarray}
The bin volumes $\Delta f \Delta \Omega$ are assumed to be small
enough that no more than one background galaxy per bin need be
considered, particularly since eventually $\Delta f \Delta \Omega \to
0$. The probability that the host galaxy is found in the $j$th bin is 
\begin{equation}
q_j = \Psi(f_j) \rho(\Omega_j) \, \Delta f \Delta \Omega \quad .
\end{equation}
Note that the occurrence of the host galaxy is {\it not} governed by
Poisson statistics since (by assumption) there is only one host
galaxy. 

Now let $H_{\rm hg}$ be the hypothesis that the error box contains the
burster's host galaxy in addition to background galaxies, while
$H_{\rm bg}$ is the hypothesis that only background galaxies are
present. First we calculate the probability $p(D \,|\, H_{\rm bg})$ of
obtaining the observed $n_d$ galaxies and not observing galaxies at
all other values of $f$ and $\Omega$ (the data proposition $D$)
assuming hypothesis $H_{\rm bg}$. This probability $p(D \,|\, H_{\rm
bg})$, the likelihood for $H_{\rm bg}$, is the product of the
probabilities of the observations in each bin. Thus 
\begin{eqnarray}
p(D \,|\, H_{\rm bg}) =& \prod_i^{n_d} p_i (n=1)\prod_{j\ne i}p_j (n=0) 
&= \prod_i^{n_d} \phi(f_i) \Delta f \Delta\Omega \prod_{{\rm all}\,j} 
   e^{-\phi(f_j) \Delta f \Delta\Omega} \\
=& e^{-\sum_{{\rm all}\, j} \phi(f_j) \Delta f \Delta\Omega}
   \prod_i^{n_d} \phi(f_i) \Delta f \Delta\Omega 
&= e^{-\int d\Omega \int_{f_{\rm lim}(\Omega)}^\infty df\, \phi(f)}
   \prod_i^{n_d} \phi(f_i) \Delta f \Delta\Omega \nonumber 
\end{eqnarray}
where in the argument of the exponential we have let $\Delta f$ and
$\Delta\Omega$ go to zero so that the sum becomes an integral.  Note
that 
\begin{equation}
\langle n_d \rangle = \int d\Omega
   \int_{f_{\rm lim}(\Omega)}^\infty df \, \phi(f)
\end{equation}
is the number of background galaxies expected to be observed within
the searched region. 

The probability $p(D \, | \, H_{\rm hg})$, the likelihood for $H_{\rm
hg}$, is 
\begin{equation}
p(D \, | \, H_{\rm hg}) = p(D \, | \, H_{\rm bg}) \left[\int d\Omega 
   \int_0^{f_{\rm lim}(\Omega)}  df \, \Psi(f)\rho(\Omega) + 
   \sum_{i=1}^{n_d} {{\Psi(f_i)\rho(\Omega_i) \, \Delta f \Delta \Omega}
   \over {\phi(f_i) \, \Delta f \Delta \Omega}}\right]
\end{equation}
where the first term in the brackets is the probability that the host
galaxy was unobservable, and the terms in the sum are the
probabilities that the $i$th observed galaxy is the host galaxy and
not a background galaxy (the denominator cancels the factor in $ p(D
\,|\, H_{\rm bg})$ that described the host galaxy as a background
galaxy). If there are galaxies in the error box, then we have to
consider the possibility that any one of them (or none of them) might
be the host galaxy; thus $p(D \,|\, H_{\rm hg})$ is a sum of the
probabilities for each possible occurrence of the host galaxy. If part
of the error box is not observed, as often occurs in searches for
contemporaneous counterparts (e.g., by GROCSE---Lee et al. 1997; Park
et al. 1997a---or LOTIS---Park et al. 1997b), then $f_{\rm lim}$ is
effectively infinite for the unobserved portion; the integral over
$\Omega$ in the first term should be over the entire error box. 

In standard ``frequentist'' statistics the ratio $ p(D \, | \, H_{\rm
hg}) / p(D \, | \, H_{\rm bg})$ can be used as a measure of how well
the presence of a host galaxy explains the data.  In Bayesian
statistics the odds ratio 
\begin{equation}
O(H_{\rm hg},H_{\rm bg}) = 
   {{p(H_{\rm hg} \,|\, D)}\over{ p(H_{\rm bg} \,|\, D)}} = 
   {{ p(H_{\rm hg})}\over{ p(H_{\rm bg})}} {{ p(D \, | \, H_{\rm hg})}
   \over{ p(D \, | \, H_{\rm bg})}}
\end{equation}
updates the ratio of the ``priors'' $p(H_{\rm hg})/ p(H_{\rm bg})$,
the probabilities that $H_{\rm hg}$ and $H_{\rm bg}$ are true based on
information available before obtaining the new data $D$, using the
``Bayes factor'' $p(D \,|\, H_{\rm hg}) / p(D \,|\, H_{\rm bg})$, the
ratio of the likelihoods.  The values of the priors $p(H_{\rm hg})$
and $p(H_{\rm bg})$ depend on our assessment of the validity of the
hypotheses, and in the absence of a strong preference for one
hypothesis over the other, it is best to set $p(H_{\rm hg})/ p(H_{\rm
bg})=1$.  A value of the odds ratio much larger than one favors the
presence of a host galaxy, while a value much less than one indicates
that no host galaxy is present; the observations cannot discriminate
between the two hypotheses for a value of order unity. Clearly the
search for host galaxies in multiple error boxes can be treated by the
product of the likelihood ratios for each error box. In our case the
ratio of the likelihoods for one error box is 
\begin{equation}
{{p(D \, | \, H_{\rm hg})} \over {p(D \, | \, H_{\rm bg}) }} =
   \int d\Omega \int_0^{f_{\rm lim}(\Omega)} df \,\Psi(f)\rho(\Omega) + 
   \sum_{i=1}^{n_d} {{\Psi(f_i)\rho(\Omega_i)}\over
   {\phi(f_i) }} \quad .
\end{equation}
The host galaxy is more likely to be present but unobserved when a
large fraction of the host galaxy flux distribution is below the
detection limit (the first term in this expression). A detected galaxy
is more likely to be the host galaxy than a background galaxy if there
is a higher probability for the host galaxy to be present at that
location and flux than a background galaxy. Note that a large value of
$\rho(\Omega_i)$ indicates a small error box.  Indeed, if the
localization probability density is set to a constant within a region
(e.g., within the 99\% contour), as is usually done, then $\rho$ would
have a value inversely proportional to the area of this region.
However, by using $\rho(\Omega)$ we allow the use of more information
about the burst localization. 
\subsection{Distribution of Host Galaxies}
The galaxy distributions required to decide whether a host galaxy is
present are based on the distribution of galaxies as a function of
flux and redshift, $\Phi(f,z)$.  This distribution is observed
directly, although cosmologists are ultimately interested in the
distribution as a function of luminosity and not flux.  The background
galaxy flux distribution used here is 
\begin{equation}
\phi(f) = \int dz \, \Phi(f,z) \quad .
\end{equation}
Although currently known imperfectly, $\Phi(f,z)$ can be established
empirically by redshift surveys; $\phi(f)$ is more easily determined 
directly from galaxy counts. 
 
We assume the host galaxy is not drawn from the same observed flux
distribution as the background galaxies.  While the background
galaxies' flux distribution is observed (e.g., by galaxy count
surveys), the host galaxy's flux distribution must be modeled.  Here
we develop the ``minimal'' cosmological model distribution; as
discussed below (\S 4), reasonable variants have been proposed which
may lead to different conclusions.  First, it is likely that a
galaxy's burst rate is proportional to the mass of the galaxy
(Fenimore et al. 1993). Most cosmological theories attribute bursts to
the mergers of two compact objects (e.g., neutron stars) within a
binary system (Eichler et al. 1989; Narayan, Paczy\'nski, \& Piran
1994); the number of such objects is presumably proportional to the
number of stars, and thus the mass. Assuming all galaxies are
characterized by the same mass-to-light ratio, the background galaxy
distribution should be weighted by the luminosity (or flux) to derive
the host galaxy distribution.  Second, the host galaxy may be modeled
to fall within a redshift range $\xi(z)$. Combining these two modeling
factors gives 
\begin{equation}
\Psi(f) = {{\int_0^\infty dz \, \xi(z) f \Phi(f,z)}\over
   {\int_0^\infty df \int_0^\infty dz \, \xi(z) f \Phi(f,z)}} \quad .
\end{equation}

The burst distance scale is unknown but has been modeled using
intensity distributions (e.g., Fenimore et al. 1993).  An intensity
quantity $G$ intrinsic to the burst (e.g., peak photon luminosity or
total energy emitted) is considered to be a fundamental burst property
(perhaps a constant); however, the normalization of $G$ is unknown.
Cosmology modifies the Euclidean $d^{-2}$ relationship between $G$ and
the related observed intensity quantity $g$ (e.g., peak photon flux or
energy fluence). Thus, the normalization of $G$ and the distance scale
are derived from the distribution of $g$ (as are other modeling
parameters).  The resulting relationship can be inverted to calculate
a range of likely distances for a given observed value of $g$.  While
most models assume that $G$ is a standard candle and that the source
density is constant per comoving volume, in the more general case the
density may evolve, $G$ may be characterized by a luminosity function,
and a K-correction may be necessary to relate $g$ and $G$ (the energy
band of the observation is redshifted).  Ultimately this modeling
should produce the distribution of sources $dN/dz$ as a function of
redshift and $g$.  Finally, 
\begin{equation}
\xi(z) = {{{dN}\over{dz}}\over{\int dz {{dN}\over{dz}}}} \quad .
\end{equation}

In most cases, a simple model is used.  Thus $\xi(z)$ is assumed to be
a delta function, providing a direct mapping between the peak photon
flux or the energy fluence and the redshift.  Given the uncertainties
of modeling the host galaxy distribution, the shape of the galaxy
luminosity function at a given redshift $\Phi(f,z)$ can be
approximated by a Schechter function (Peebles 1993, p.~120),
\begin{equation}
\psi(y) = \psi_0 y^\alpha e^{-y} \quad ,
\end{equation}
where $y=L/L_*=f/f_*$.  The intensity scale, $L_*$ or $f_*$, is
typically measured as the absolute magnitude in a given spectral band.
Note that a K-correction should be applied for redshifts more than a
few tenths (the flux observed in a given energy band was emitted by
the galaxy in a different band); in addition, the luminosity function
underwent evolution both in scale (i.e., the value of $L_*$) and
normalization.  Recent surveys find: for the b$_{\rm j}$ band $M_{{\rm
b}_{\rm j}}=-19.72 \pm 0.09$ and $\alpha = -1.14 \pm 0.08$ (Ratcliffe
et al. 1997); 
%
%
for the K-band $M_K = -23.12$ and $\alpha = -0.91$ (Gardner et al.
1997); 
%
%
and for the R-band $M_R = -20.29 \pm 0.02$ and $\alpha = -0.70 \pm
0.05$ (Lin et al. 1996).   
%
%
To all these expressions for $M_*$ must be added an additional term
$5\log h$, where $h=H_0$/(100 km s$^{-1}$ Mpc$^{-1}$), resulting from
the uncertainty in Hubble's Constant $H_0$; however, this dependence
on the value of $H_0$ is cancelled by the $H_0$ dependence in the
relationship between the flux and $z$ ($M_*$ is derived from
observations of magnitude vs. redshift).  When we do need $H_0$, we
use $h=0.75$. Since $\alpha$ is of order -1, we approximate $\Psi(f)$
as a simple exponential: 
\begin{equation}
\Psi(f)=\exp[-f/f_*]/f_* \quad .
\end{equation}

Using the approximation for small values of $z$,
\begin{equation}
m_* = M_* +5\log[3\times10^8 z] \quad \hbox{and} \quad
f_* (z) = f_0 10^{-0.4 M_*} [3\times10^8 z]^{-2} \ , 
\end{equation}
where $f_0$ is the normalizing flux (i.e., the flux of a 0 magnitude
object) for a given band, and $m_*$ is the apparent magnitude
corresponding to $f_*$.  
\subsection{Sensitivity}
We can evaluate how well our methodology discriminates between the
presence or absence of a host galaxy in a given error box.  Since
$\phi(f)$ and $\Psi(f)\rho(\Omega)$ are the probabilities of the
presence of a background galaxy and the host galaxy, respectively, in
a given patch of sky $\Omega$ at a flux $f$, the expected value of the
Bayes factor is 
\begin{eqnarray}
\left\langle{{p(D \, | \, H_{\rm hg})} \over 
   {p(D \, | \, H_{\rm bg}) }} \right\rangle &=&
   \int d\Omega \int_0^{f_{\rm lim}(\Omega)} df \,\Psi(f)\rho(\Omega) 
   \nonumber \\ &+& 
   \int d\Omega \int_{f_{\rm lim}(\Omega)}^\infty df \, 
   [\phi(f)+\Psi(f)\rho(\Omega)]
   {{\Psi(f)\rho(\Omega)}\over{\phi(f) }} \nonumber \\
   &=& 1+\int d\Omega \int_{f_{\rm lim}(\Omega)}^\infty df \, 
   {{\Psi(f)^2\rho(\Omega)^2}\over{\phi(f) }}
\end{eqnarray}
if a host galaxy is present.  If there is no host galaxy then both the
second term in the brackets (i.e., $\Psi(f)\rho(\Omega)$) in the first
equation of eq.~(14) and the integral in the second equation should
not be included.  Thus on average $\langle p(D \, | \, H_{\rm hg}) /
p(D \, | \, H_{\rm bg})\rangle =1$ without a host galaxy, which may
seem surprising, but results from background galaxies being
occasionally mistaken for the host galaxy. Under the hypothesis
$H_{\rm bg}$ that there are no host galaxies, $p(D \, | \, H_{\rm hg})
/ p(D \, | \, H_{\rm bg})$ will be less than 1 in most error boxes.
However, in those cases where a background galaxy with the flux
expected for the host galaxy falls in the error box, $p(D \, | \,
H_{\rm hg}) / p(D \, | \, H_{\rm bg})$ will be greater than 1.  The
more unlikely such an occurrence, the smaller the value of $\phi$ and
therefore the larger the value of the second term of $p(D \, | \,
H_{\rm hg}) / p(D \, | \, H_{\rm bg})$ when a background galaxy is
mistaken for the host galaxy. 

The methodology's power to determine that a host galaxy is present in
an error box for a given observation depends on the value of $\int
d\Omega \int_{f_{\rm lim} (\Omega)}^\infty df \, \Psi(f)^2
\rho(\Omega)^2 / \phi(f)$, the term added by the presence of a host
galaxy.  To evaluate this expression we assume that a constant
probability error box (i.e., $\rho=1/\Omega_0$ over an area
$\Omega_0$) is searched to a uniform detection threshold $f_{\rm
lim}$. We approximate the cumulative galaxy count distribution over
the error box as $N(>f)=(f/f_b)^{-3/2}$ (the distribution expected for
a constant galaxy density in three-dimensional Euclidean space), where
$f_b$ is the flux at which one background galaxy is expected in the
error box. Since $N(>f)\propto \Omega_0$, $f_b$ will vary from error
box to error box as $f_b\propto \Omega_0^{2/3}$.  From $N(>f)$ we
derive $\phi(f)= (3/2 f_b\Omega_0) (f/f_b)^{-5/2}$.  As discussed in
\S 2.2, we use a weighted Schechter function for the host galaxy
distribution function, $\Psi(f)=\exp[-f/f_*]/f_*$.  Consequently if a
host galaxy is present 
\begin{eqnarray}
\left \langle {{p(D \, | \, H_{\rm hg})}\over 
   {p(D \, | \, H_{\rm bg})}} \right\rangle
   &=& 1+{1\over{2^{5/2}3}} \left({{f_*}\over{f_b}}\right)^{3/2}
   \int_{2f_{\rm lim}/f_*}^\infty du \, u^{5/2} e^{-u} \nonumber \\
   &=& 1+{1\over{2^{5/2}3}} \left({{f_*}\over{f_b}}\right)^{3/2}
   \Gamma \left({7\over2},2f_{\rm lim}/f_*\right)
\end{eqnarray}
where $\Gamma(a,x)$ is the incomplete gamma function. Since most of
the area under the curve $ u^{5/2} e^{-u}$ is above $u=1$, the value
of $\Gamma \left(7/2,2f_{\rm lim}/f_*\right)$, and therefore of
$\langle p(D \, | \, H_{\rm hg}) / p(D \, | \, H_{\rm bg}) \rangle$ is
relatively insensitive to $ f_{\rm lim}$ as long as $f_{\rm
lim}/f_*<1/2$, that is, when the detection threshold is less than the
expected host galaxy flux.  On the other hand, $\langle p(D \, | \,
H_{\rm hg}) / p(D \, | \, H_{\rm bg}) \rangle$ is very sensitive to
the value of $f_b$, the flux at which we expect one background galaxy
in the error box.  Since $\Gamma(7/2)=3.32335$, $\langle p(D \,|\,
H_{\rm hg})/ p(D \,|\, H_{\rm bg}) \rangle \simeq 1 +
0.1958(f_*/f_b)^{3/2}$.  Our methodology (and most likely any
methodology) can discriminate between the presence and absence of a
host galaxy in given error box when, on average, the host galaxy is
expected to be much brighter than the brightest background galaxy. 

These results were derived analytically by integrating over the number
of galaxies detected in the error box.  This integration treats the
number of galaxies as a continuous quantity.  We have verified the
results with simulations with discrete bursts made under the same
assumptions that led to eq.~(15).  Define $\alpha=f_{\rm lim}/f_*$ and
$\beta = f_b/f_*$.  Then our analytic formalism gives 
\begin{eqnarray}
\langle n_h \rangle &=& \exp(-\alpha) \ , \quad
   \langle n_b \rangle = (\beta/\alpha)^{3/2} \ , \quad
   \langle O_b \rangle = 1 \ , \nonumber \\
\langle O_h \rangle &=& 1+0.1958 \beta^{-3/2} 
   {{\Gamma(7/2,2\alpha)}\over{3.32335}} \ ,
\end{eqnarray} 
where $n_h$ and $n_b$ are the numbers of host galaxies and background
galaxies, respectively, detected in an error box. $\langle O_h
\rangle$ ($\langle O_b \rangle$) is the odds ratio if host and
background galaxies (only background galaxies) are present, assuming
the prior ratio is 1.  In each trial of a given simulation we
generated one host galaxy with normalized flux $\epsilon_h=f_h/f_*$
and 100 background galaxies with normalized fluxes
$\epsilon_i=f_i/f_*$, where all the $\epsilon$ values were drawn from
the appropriate distribution functions.  Then 
\begin{equation}
\langle O_b \rangle = 1-e^{-\alpha} + \sum_{i=1}^{100} 
   {{e^{-\epsilon_i} \epsilon_i^{5/2}} \over
   {3/2 \, \beta^{3/2}}} \theta(\epsilon_i-\alpha) 
\quad \hbox{and} \quad
\langle O_h \rangle = \langle O_b \rangle + 
   {{e^{-\epsilon_h} \epsilon_h^{5/2}} \over
   {3/2 \, \beta^{3/2}}} \theta(\epsilon_h-\alpha) \quad ,
\end{equation}
where $\theta(x)$ is the Heaviside function which is used here to
enforce the requirement that a galaxy be detectable to be included. 
The results of our simulations are provided by Table~1.  As can be
seen, the agreement with the analytic formulae in eq.~(16) is very
good.  The largest deviations occur when $\langle n_b \rangle$ is very
small, and even $10^4$ trials do not provide sufficient statistics. 
\section{APPLICATION}
In this section we apply our methodology to observations of gamma-ray
burst error boxes reported in the literature.  Not all details are
provided in these publications, and we make a number of simplifying
assumptions; for example, a constant probability density over the
error box is used. We write the ratio of likelihoods (eq.~[7]) as $p(D
\, | \, H_{\rm hg}) / p(D \, | \, H_{\rm bg}) = q_1 + \sum_{i=1}^{n_d}
q_{2i}$ where 
\begin{equation}
q_1 = \int d\Omega \int_0^{f_{\rm lim}(\Omega)} df \,\Psi(f)\rho(\Omega) 
   = 1 - {{\Omega_s}\over{\Omega_0}} e^{-f_{\rm lim}/f_*}
\end{equation}
and
\begin{equation}
q_{2i} = {{\Psi(f_i)\rho(\Omega_i)}\over {\phi(f_i) }} 
   = {{e^{-f_i/f_*}}\over{f_* \Omega_0 \phi(f_i) }} \quad .
\end{equation}
In these expressions $\Omega_0$ is the size of the error box ($\rho
\propto 1/\Omega_0$) and $\Omega_s\le\Omega_0$ is the size of the
searched region.  Since we are testing the ``minimal'' cosmological
model, for the host galaxy distribution function $\Psi(f)$ we use the
weighted Schechter function in eq.~(12). We derived the distribution
of background galaxies $\phi(f)$ from the review paper by Koo \& Kron
(1992).  In their Figure~1 they compile galaxy counts in the K, R and
b$_{\rm j}$ bands from a number of different sources.  We
parameterized the differential galaxy count distribution by choosing
values which fell within the cluster of observed data points. 
\subsection{K-Band Observations of Larson \& McLean (1997)}
Larson \& McLean (1997) observed the error boxes of 9 recent bursts
localized by the third Interplanetary Network as well as a number of
control fields, primarily in the K band.  Many of the error boxes were
only partially observed because of corrections to the error boxes
after the observations.  They found many relatively bright galaxies
both within and immediately outside of the error boxes. The error
boxes are typically $\sim8$ arcmin$^2$, and background galaxies are
expected; Larson \& McLean recognize that they cannot distinguish
between host and background galaxies in their data.  Based on their
observations of control fields and galaxy counts from the literature,
they report that the overall galaxy density in and near the error
boxes is about a factor of 2 greater larger than average, which might
result from clustering around the host galaxy. 

Although Larson \& McLean recognize that their observations are
consistent with, but do not prove, the existence of host galaxies,
their earlier work (Larson, McLean \& Becklin 1996) was interpreted as
showing there was not a host galaxy issue.  Therefore we analyzed
their observations to determine what statement could be made about the
existence of host galaxies.  In Table~2 we list for each of the nine
error boxes studied by Larson \& McLean the size of the box, the
fraction of the box covered by the observations, and the K magnitude
of the brightest galaxy in the box. We assume the probability density
$\rho$ is constant within the error boxes.  Larson (1997) provides an
estimated redshift for these bursts.  Ideally, we would use a list of
magnitudes for all detected galaxies above the limiting magnitude
(here $K\sim 18.5$), but here we only have the brightest galaxy. 
Therefore, we use the magnitude of the brightest galaxy as both the
limiting magnitude and the magnitude of the single detected galaxy for
each box. 

The flux at which we expect to find background galaxies is $f_b =5.8
\times 10^{-28} \Omega_0^{2/3} \sim 2.32 \times 10^{-27}$
erg~s$^{-1}$~cm$^{-2}$~Hz$^{-1}$ for $\Omega_0\sim 8$~arcmin$^{2}$.
Therefore $f_b$, $f_*$, and the flux of the detection threshold (which
is also the flux of the brightest galaxy) are all comparable. In
Table~2 we compute the various terms of the likelihood ratio.  The
first term, $q_1$, which is also the probability that the host galaxy
is fainter than the detection threshold, increases when the box is not
observed completely.  The product of the likelihood ratios $q_{\rm
tot}$ for each error box is the likelihood ratio for the ensemble;
here the product is 0.248, which indicates that we cannot determine
whether the expected host galaxies are present or absent. 
%
%
\subsection{{\it HST} Observations of Schaefer et al. (1997)}
Schaefer et al. (1997) observed four small error boxes with the WFPC2
on {\it HST} in both the B and UV bands (they also studied a much
larger error box with the FOC; this error box is not considered here).
Here we apply our methodology to their B-band observations; in our
analysis we use b$_{\rm j}$ distribution functions. In each case a
detection threshold is given.  Sources are found in three of the four
error boxes, but we consider only the small number of sources which
are identified as galaxies. 

Two of the error boxes were only partially observed.  Schaefer et al.
stated that ground-based observations of the GRB~790613 error box show
there are no sources {\it HST} would have detected in the 10\% of the
error box which was not observed.  Similarly, Schaefer (1997, private
communication) reports there are no galaxies to the {\it HST}
detection limit in the $\sim15$\% of the error box unobserved by {\it
HST}.  Thus $\Omega_s=\Omega_0$ (see eq.~[18]) for these two boxes.

Note that Larson (1997) and Schaefer et al. (1997) give $z$ values for
GRB~920406 which differ by a factor of two.  Larson uses redshifts
provided by E.~Fenimore, while Schaefer et al. give distances
calculated from peak energy fluxes assuming a peak luminosity of
$6\times10^{50}$~ergs~s$^{-1}$.  This peak luminosity was calculated
for $H_0=75$~km~s$^{-1}$~Mpc$^{-1}$, which we used to convert the
distances to redshifts. 

Table~3 presents the results of applying our methodology to the {\it
HST} observations. As Schaefer et al. concluded, the detection
thresholds are sufficiently fainter than the expected $f_*$ such that
the host galaxy should have been observed.  In addition the observed
galaxies are as faint as the expected background galaxies for such
small error boxes; for a threshold of B=23 we expect $\sim 3$ galaxies
per arcmin$^{2}$.  The product of the likelihood ratios is
$2\times10^{-6}$, which indicates that these observations strongly
favor the hypothesis that only background galaxies and no host
galaxies are found in the error boxes observed by {\it HST}.  Note
that $f_b = 7.7\times10^{-29} \Omega_0^{2/3}$ erg s$^{-1}$ cm$^{-2}$
Hz$^{-1}$ and thus $f_*\gg f_b$; by \S 2.3 our methodology should be
able to determine that host galaxies are present in these error boxes.
\section{DISCUSSION}
Almost all the quantities required for analyzing an error box are
observables, although they may be difficult to derive and may be
characterized by large uncertainties.  However, the host galaxy
distribution function is model dependent, and any conclusions based on
applying our methodology is a statement about the validity of the host
galaxy model.  Here we assume that bursts occur within galaxies, and
that the number of sources is proportional to the luminosity of the
galaxy. This model is consistent with the scenario where bursts result
from the merger of neutron star binaries.  In analyzing the
observations from the literature we used the ``minimal'' cosmological
model, although in \S 2.2 we outline how a distribution for the source
redshift can be derived. 

As discussed above, this methodology can also be applied to scenarios
where the burst source is ejected from the host galaxy by expanding
the error box by the angular distance the source may have traveled
before bursting. For arcmin$^2$ scale error boxes, broadening the
error box may be inconsequential, but for the well-localized bursts
resulting from the newly-discovered optical transients, the size of
the error box may be determined primarily by the distance the source
traveled before bursting. 

While the minimal cosmological host galaxy model we use is consistent
with current modeling assumptions in studies of the burst database,
reasonable variants can result in significantly different conclusions,
and the analysis of an error box may have to be revisited as the
source models evolve.  For example, the observed burst distribution
may be characterized by a broad luminosity function (Horack et al.
1994, 1996 and Hakkila et al. 1995, 1996 found that the luminosity
function must be narrow, but Loredo and Wasserman 1997a,b and Brainerd
1997 dispute this conclusion), allowing a given burst to originate
over a broad redshift band.  The source density may be greater in
galaxies which have undergone starbursts, which may favor small
galaxies at moderate redshifts (Sahu et al. 1997). 

Our methodology compares two hypotheses: 1) a host galaxy is present
in addition to background galaxies; or 2) only background galaxies are
present.  This analysis is applied to each error box independently;
the product of the likelihood ratios compares these hypotheses for the
data set as a whole.  This methodology does not test directly whether
the candidate host galaxies are indeed distributed according to the
model distribution function.  Fenimore et al. (1993) introduced a test
which considers the distribution of the statistic $S=\int_0^{f_i}
df\,\Psi(f)$, where $f_i$ is the flux of the brightest galaxy in, or
the detection threshold for, the $i$th error box. If the $f_i$ indeed
correspond to the host galaxies, then the statistic $S$ should be
distributed uniformly between 0 and 1, and should have an average
value of 1/2. However, this test does not evaluate whether $f_i$
corresponds to the host galaxy. 
\section{SUMMARY}
We have developed a methodology to evaluate whether the detections and
nondetections of galaxies in gamma-ray burst error boxes are
consistent with the presence of the burst source's host galaxy, or
whether all the detections can be attributed to unrelated background
galaxies.  This methodology relies on the distribution of background
galaxies, which is observed, and the flux distribution for the host
galaxy, which must be modeled.  Any conclusions are dependent on the
host galaxy model.  In addition to evaluating the observations of a
particular error box, our methodology also predicts its likely
sensitivity for a given error box.  Of course, the methodology can be 
used for other candidate host sources, or for similar astrophysical 
problems.

The methodology allows the maximal use of observational data.  Thus a
variable probability density for the burst's location can be used
instead of assuming that this probability is constant across the error
box.  The detection threshold is allowed to vary across the error box;
this permits the treatment of partially observed error boxes. 

As examples we applied this methodology to the K-band observations of
Larson \& McLean (1997) and the {\it HST} observations of Schaefer et
al. (1997). Larson \& McLean found a large number of galaxies in or
near 9 error boxes, but our analysis shows that these detections can
easily be explained as background galaxies. We find that the error
boxes Larson et al. observed were too large to discriminate between
the presence or absence of a host galaxy. Schaefer et al. (1997) found
only faint galaxies; our analysis shows that brighter host galaxies
should have been observed, but were not. Our sensitivity analysis
verifies that these error boxes are small enough to determine that a
host galaxy is indeed present. Therefore, within the assumptions of
the ``minimal'' cosmological model used in our analysis (e.g.,
standard candle bursts; no evolution; the number of potential sources
is proportional to the size of the galaxy), the expected host galaxies
are absent. 

In the future we will apply this methodology to other observations of
small error boxes, and explore other host models. 

\acknowledgments
We thank F.~Vrba and C.~Luginbuhl for stimulating discussions.  We
appreciate the comments and encouragement of the referee, B.~Schaefer.
D.~Band's gamma-ray burst research is supported by the {\it CGRO}
guest investigator program and NASA contract NAS8-36081.  D.~Hartmann
acknowledges support from the {\it CGRO} guest investigator program.

\clearpage

\begin{deluxetable}{l l r c c c c c c c c}
\tablecolumns{11}
\scriptsize
\tablewidth{0pc}
\tablecaption{Sensitivity Simulations}
\tablehead{
\colhead{$\alpha$\tablenotemark{a}} & 
\colhead{$\beta$\tablenotemark{b}} & 
\colhead{$N_{\rm trial}$\tablenotemark{c}} & 
\multispan{2}{\qquad\qquad$\langle n_h \rangle$\tablenotemark{d}} & 
\multispan{2}{\qquad\qquad$\langle n_b \rangle$\tablenotemark{e}} & 
\multispan{2}{\qquad\qquad$\langle O_b \rangle$\tablenotemark{f}} & 
\multispan{2}{\qquad\qquad$\langle O_h \rangle$\tablenotemark{g}} \\
&&& 
\colhead{\quad Sim.\tablenotemark{h}} & 
\colhead{Calc.\tablenotemark{i}} &
\colhead{\quad Sim.\tablenotemark{h}} & 
\colhead{Calc.\tablenotemark{i}} &
\colhead{\quad Sim.\tablenotemark{h}} & 
\colhead{Calc.\tablenotemark{i}} &
\colhead{\quad Sim.\tablenotemark{h}} & 
\colhead{Calc.\tablenotemark{i}} 
}
\startdata
1.    & 1.    & $10^4$ & 0.3677 & 0.3679 & 1.0065 & 1.0000 & 1.0014 
& 1. & 1.1537 & 0.9325 \\
0.1   & 1.    & $10^4$ & 0.9030 & 0.9048 & 31.610 & 31.623 & 0.9908 
& 1. & 1.1872 & 1.1955 \\
1.0   & 0.1   & $10^4$ & 0.3611 & 0.3679 & 0.0298 & 0.0316 & 0.9977 
& 1. & 5.7152 & 5.6079 \\
0.1   & 0.1   & $10^4$ & 0.8985 & 0.9048 & 1.0016 & 1.0000 & 1.0269 
& 1. & 7.1515 & 7.1901 \\
0.01  & 0.1   & $10^4$ & 0.9888 & 0.9900 & 31.558 & 31.623 & 0.9982 
& 1. & 7.1540 & 7.1917 \\
1.    & 0.01  & $10^4$ & 0.3618 & 0.3679 & 0.0010 & 0.0010 & 0.9923 
& 1. & 150.43 & 153.48 \\
0.1   & 0.01  & $10^4$ & 0.9053 & 0.9048 & 0.0334 & 0.0316 & 0.8232 
& 1. & 196.68 & 196.78 \\
0.01  & 0.01  & $10^4$ & 0.9897 & 0.9900 & 1.0109 & 1.0000 & 0.8119 
& 1. & 195.91 & 196.83 \\
0.001 & 0.01  & $10^4$ & 0.9987 & 0.9990 & 31.627 & 31.623 & 0.9087 
& 1. & 196.88 & 196.83 \\
1.    & 0.001 & $10^4$ & 0.3718 & 0.3679 & 0.0000 & 0.0000 & 0.6321 
& 1. & 4901.4 & 4829.7 \\
0.1   & 0.001 & $3\times 10^4$ & 0.9079 & 0.9048 & 0.0011 & 0.0010 
& 0.7906 & 1. & 6186.0 & 6192.1 \\
0.01  & 0.001 & $3\times 10^4$ & 0.9905 & 0.9900 & 0.0034 & 0.0316 
& 1.1030 & 1. & 6191.3 & 6193.7 \\
0.001 & 0.001 & $10^4$ & 0.9997 & 0.9990 & 1.0064 & 1.0000 & 1.0062 
& 1. & 6203.2 & 6193.7 \\
$10^{-4}$ & 0.001 & $10^4$ & 0.9998 & 0.9999 & 31.654 & 31.623 & 1.0161 
& 1. & 6135.5 & 6193.7
\enddata
\tablenotetext{a}{The ratio $f_{\rm lim}/f_*$.}
\tablenotetext{b}{The ratio $f_b/f_*$.}
\tablenotetext{c}{The number of trials in the simulation.}
\tablenotetext{d}{The fraction of the trials in which the host galaxy 
is detected.}
\tablenotetext{e}{The average number of background galaxies detected 
per error box.}
\tablenotetext{f}{The average value of the odds ratio when only background 
galaxies are present, and the ratio of priors is set to unity.}
\tablenotetext{g}{The average value of the odds ratio when the host 
galaxy is present in addition to background 
galaxies.  The ratio of priors is set to unity.}
\tablenotetext{h}{Results of simulation.}
\tablenotetext{i}{Calculated using eq.~(16).}
\end{deluxetable}

\begin{deluxetable}{l c c c c c c c c c c}
\tablecolumns{11}
\scriptsize
\tablewidth{0pc}
\tablecaption{Analysis of the Larson \& McLean (1997) Observations}
\tablehead{
\colhead{Burst} & 
\colhead{$z$\tablenotemark{a}} & 
\colhead{$\Omega_0$\tablenotemark{b}} & 
\colhead{Cov.\tablenotemark{c}} &
\colhead{K$_i$} & 
\colhead{$f($K$_i)$\tablenotemark{d}} & 
\colhead{$f_*(z)$\tablenotemark{e}} & 
\colhead{$\phi($K$_i)$\tablenotemark{f}} & 
\colhead{$q_1$\tablenotemark{g}} & 
\colhead{$q_{2i}$\tablenotemark{h}} & 
\colhead{$q_{\rm tot}$\tablenotemark{i}}
}
\startdata
GRB~910122 &
0.19 & 19.33& 0.41 & 15.3 & $4.70\times10^{-27}$ & $3.38\times10^{-27}$ &
 $1.87\times10^{+25}$ & 0.898 & 0.203 & 1.102 \nl
GRB~910219 &
0.20 &  7.30& 1.0  & 15.7 & $3.25\times10^{-27}$ & $3.05\times10^{-27}$ &
 $4.49\times10^{+25}$ & 0.656 & 0.344 & 1.000 \nl
GRB~920325 &
0.22 & 26.5 & 0.87 & 14.0 & $1.56\times10^{-26}$ & $2.52\times10^{-27}$ &
 $1.22\times10^{+24}$ & 0.998 & 0.025 & 1.024 \nl
GRB~920406 &
0.12 &  1.65& 1.0  & 17.1 & $8.96\times10^{-28}$ & $8.47\times10^{-27}$ &
 $8.26\times10^{+26}$ & 0.100 & 0.078 & 0.178 \nl
GRB~920501 &
0.17 &  2.60& 0.91 & 14.8 & $7.45\times10^{-27}$ & $4.22\times10^{-27}$ &
 $6.41\times10^{+24}$ & 0.844 & 2.429 & 3.274 \nl
GRB~920525 &
0.15 &  3.54& 1.0  & 17.0 & $9.83\times10^{-28}$ & $5.42\times10^{-27}$ &
 $6.87\times10^{+26}$ & 0.166 & 0.063 & 0.229 \nl
GRB~920711 &
0.23 &  2.50& 0.93 & 15.5 & $3.91\times10^{-27}$ & $2.31\times10^{-27}$ &
 $2.90\times10^{+25}$ & 0.830 & 1.097 & 1.927 \nl
GRB~920720 &
0.20 &  3.78& 0.57 & 15.5 & $3.91\times10^{-27}$ & $3.05\times10^{-27}$ &
 $2.90\times10^{+25}$ & 0.842 & 0.830 & 1.672 \nl
GRB~920723 &
0.11 &  4.46& 0.88 & 16.0 & $2.47\times10^{-27}$ & $1.01\times10^{-26}$ &
 $8.65\times10^{+25}$ & 0.311 & 0.201 & 0.512 
\enddata
\tablenotetext{a}{Burst redshift from Larson (1997).}
\tablenotetext{b}{Size of the error box in arcmin$^2$.}
\tablenotetext{c}{Fraction of the error box covered by the 
observations.}
\tablenotetext{d}{The flux corresponding to a given K magnitude,
using a K=0 flux of $0.62 \times 10^{-20}$ erg s$^{-1}$ cm$^{-2}$
Hz$^{-1}$.} 
\tablenotetext{e}{The K-band flux (erg s$^{-1}$ cm$^{-2}$ Hz$^{-1}$)
corresponding to an L$_*$ galaxy at a given redshift based on Gardner
et al. (1997).} 
\tablenotetext{f}{The differential galaxy distribution (galaxies 
arcmin$^{-2}$ flux$^{-1}$) based on the cumulative distribution in 
Figure~1 of Koo \& Kron (1992).}
\tablenotetext{g}{The probability that the host galaxy is fainter than
the detection limit.} 
\tablenotetext{h}{The value of $q_{2i}$ for the detected galaxy.}
\tablenotetext{i}{Total likelihood ratio for the error box, the sum of
$q_1$ and all the $q_{2i}$.} 
\end{deluxetable}

\begin{deluxetable}{l c c c c c c c c c c c}
\tablecolumns{12}
\scriptsize
\tablewidth{0pc}
\tablecaption{Analysis of the Schaefer et al. (1997) Observations}
\tablehead{
\colhead{Burst} & \colhead{$\Omega_0$\tablenotemark{a}} & 
\colhead{$z$\tablenotemark{b}} & \colhead{$f_*(z)$\tablenotemark{c}} & 
\colhead{B$_{\rm lim}$\tablenotemark{d}} & 
\colhead{$f(B_{\rm lim})$\tablenotemark{e}} & \colhead{B$_i$} &
\colhead{$f(B_i)$\tablenotemark{e}} & 
\colhead{$\phi(B_i)$\tablenotemark{f}} &
\colhead{$q_1$\tablenotemark{g}} & 
\colhead{$q_{2i}$\tablenotemark{h}} & 
\colhead{$q_{\rm tot}$\tablenotemark{i}}
}
\startdata
GRB~790325 &  2.0 & 0.145 & $1.81\times10^{-27}$ & 23.0  & 
   $2.80\times10^{-29}$ & 22.53 & $4.32\times10^{-29}$ & 
   $4.53\times10^{28}$ & 0.015 & 0.006 & 0.021 \nl
GRB~790406 & 0.26 & 0.115 & $2.88\times10^{-27}$ & 22.8  & 
   $3.37\times10^{-29}$ & ---   & ---                  & 
   ---                 & 0.012 & ---    & 0.012 \nl
GRB~790613 & 0.76 & 0.100 & $3.81\times10^{-27}$ & 23.2  & 
   $2.33\times10^{-29}$ & 21.29 & $1.35\times10^{-28}$ & 
   $4.61\times10^{27}$ & 0.006 & 0.072 &        \nl
           &      &       &                      &       &
                        & 21.61 & $1.01\times10^{-28}$ & 
   $8.32\times10^{27}$ &        & 0.040 &        \nl
           &      &       &                      &       &
                        & 21.57 & $1.05\times10^{-28}$ & 
   $7.73\times10^{27}$ &        & 0.044 & 0.162 \nl
GRB~920406 &  2.0 & 0.263 & $5.53\times10^{-28}$ & 23.0  & 
   $2.80\times10^{-29}$ & ---   & ---                  &  
  ---                & 0.049 &        & 0.049  \nl
\enddata
\tablenotetext{a}{Size of the error box in arcmin$^2$.}
\tablenotetext{b}{Burst redshift converted (assuming $h=0.75$) from
distance given by Schaefer et al., which in turn is based on the peak
flux.} 
\tablenotetext{c}{The B-band flux (erg s$^{-1}$ cm$^{-2}$ Hz$^{-1}$)
corresponding to an L$_*$ galaxy at a given redshift based on Ratcliffe
et al. (1997).} 
\tablenotetext{d}{The limiting B-magnitude.}
\tablenotetext{e}{The flux corresponding to a given B magnitude, using
a B=0 flux of $4.44 \times 10^{-20}$ erg s$^{-1}$ cm$^{-2}$
Hz$^{-1}$.} 
\tablenotetext{f}{The differential galaxy distribution (galaxies 
arcmin$^{-2}$ flux$^{-1}$) based on the cumulative distribution in 
Figure~1 of Koo \& Kron (1992).}
\tablenotetext{g}{The probability that the host galaxy is fainter than
the detection limit.  The value is reported on the first line for a
given error box.} 
\tablenotetext{h}{The value of $q_{2i}$ for the detected galaxy.}
\tablenotetext{i}{Total likelihood ratio for the error box, the sum of
$q_1$ and all the $q_{2i}$.  The resulting value is reported on the
last line for a given error box.} 
\end{deluxetable}


\end{document}